# Asynchronous Integration of Real-Time Simulators for HIL-based Validation of Smart Grids

Catalin Gavriluta, Georg Lauss, Thomas I. Strasser
AIT Austrian Institute of Technology
Vienna, Austria
firstname.lastname@ait.ac.at

Juan Montoya, Ron Brandl
Fraunhofer Institute for Energy Economics and Energy System Technology (IEE)
Kassel, Germany
firstname.lastname@iee.fraunhofer.de

Panos Kotsampopoulos
National Technical University of Athens
Athens, Greece
kotsa@power.ece.ntua.gr

***Abstract*—As the landscape of devices that interact with the electrical grid expands, also the complexity of the scenarios that arise from these interactions increases. Validation methods and tools are typically domain specific and are designed to approach mainly component level testing. For this kind of applications, software and hardware-in-the-loop based simulations as well as lab experiments are all tools that allow testing with different degrees of accuracy at various stages in the development life-cycle. However, things are vastly different when analysing the tools and the methodology available for performing system-level validation. Until now there are no available well-defined approaches for testing complex use cases involving components from different domains. Smart grid applications would typically include a relatively large number of physical devices, software components, as well as communication technology, all working hand in hand. This paper explores the possibilities that are opened in terms of testing by the integration of a real-time simulator into co-simulation environments. Three practical implementations of such systems together with performance metrics are discussed. Two control-related examples are selected in order to show the capabilities of the proposed approach.**

## I. Introduction

Driven by the goals of efficiency, reliability, and renewability, the power system is transitioning towards a smart grid [1], [2]. Distributed generation and energy storage, electric vehicles, power electronics, smart meters, and Phasor Measurement Units (PMU) are just some of the new game changers in the field.

As the landscape of devices that interact with the electrical grid expands, also the complexity of the scenarios that arise from these interactions increases. Consequently, numerous advanced control strategies have started to migrate from the control theory to the smart grid literature. Numerous multi-agent and consensus-based control strategies have been proposed in the last few years, addressing various operational challenges of future grids, see [3], [4]. Similarly, topics related to distributed control and optimisation have also started to become more popular in the electrical engineering literature, see [5]–[7].

One aspect that all of these approaches have in common is that they all involve a large numbers of distributed physical devices controlled by various software components and advanced algorithms, all linked together by communication technology. With the current movement towards the decentralisation of the energy supply and the advances from the domain of the "Internet of Things" (IoT), it is expected that most smart grid applications will fit this description. However, before any of them has a chance of becoming a reality, proper tools and methodologies need to be made available to the engineers in order to extensively test and validate them, see [8].

The goal of this paper is to explore the possibilities that are opened in terms of testing and validating advanced control strategies and smart grid applications by the integration of a real-time simulator into co-simulation environments.

The paper starts with Section II by briefly outlining existing simulation-based approaches that allow the integration of real-time simulators. Then, in Section III three practical implementations of such an integration are presented together with performance results, followed by Section IV which showcases two applications which were evaluated using the proposed approaches. The first one tackles the validation of a distributed optimal power flow algorithm in a dc microgrid, whereas the second one deals with a coordinated optimal voltage control strategy for low-voltage distribution feeders. Finally, Section V concludes the paper.

## II. Simulation-based Validation Methods

As outlined above there are no off-the-shelf tools available for testing complex smart grid applications that involve components from different domains. However, several approaches can be found in the literature, spanning from lab experiments in [3] to custom-built co-simulation platforms in [9].

While suitable for testing small-scale systems, laboratory prototypes do not scale well with the size of the system. Meanwhile, co-simulation platforms, when properly designed, allow for a larger degree of flexibility and scalability. Such a co-simulation framework is mosaik, see [10], a Python-based open source software package. Mosaik allows the coupling of existing simulators and models in order to tackle complex smart-grid applications.

While the mosaik framework is mainly focused on coupling of offline simulators, the approach proposed in [11], envisions a Simulation Message Bus (SMB) for co-simulation and rapid prototyping of networked systems. This approach would enable the coupling of both offline software modules and real-time digital simulators.

The generic architecture of a SMB-based co-simulation is shown in Fig. 1. The main component is the simulation databus. Input and output interfaces wrap around the core and act as a middle layer that allows data to be injected or extracted from the message bus. Depending on the sample rate at which data needs to be exchanged with the core, specifically designed task processing units will be needed for the purpose of the respective application. These task processing units represent functional units implemented in software and are labelled as $IN_{O1}$, $IN_{O2}$, $IN_{O3}$, . . . $IN_{ON}$ in Fig. 1. Their function is to design custom software or hardware adaptations for each application or simulator that participates in the co-simulation.

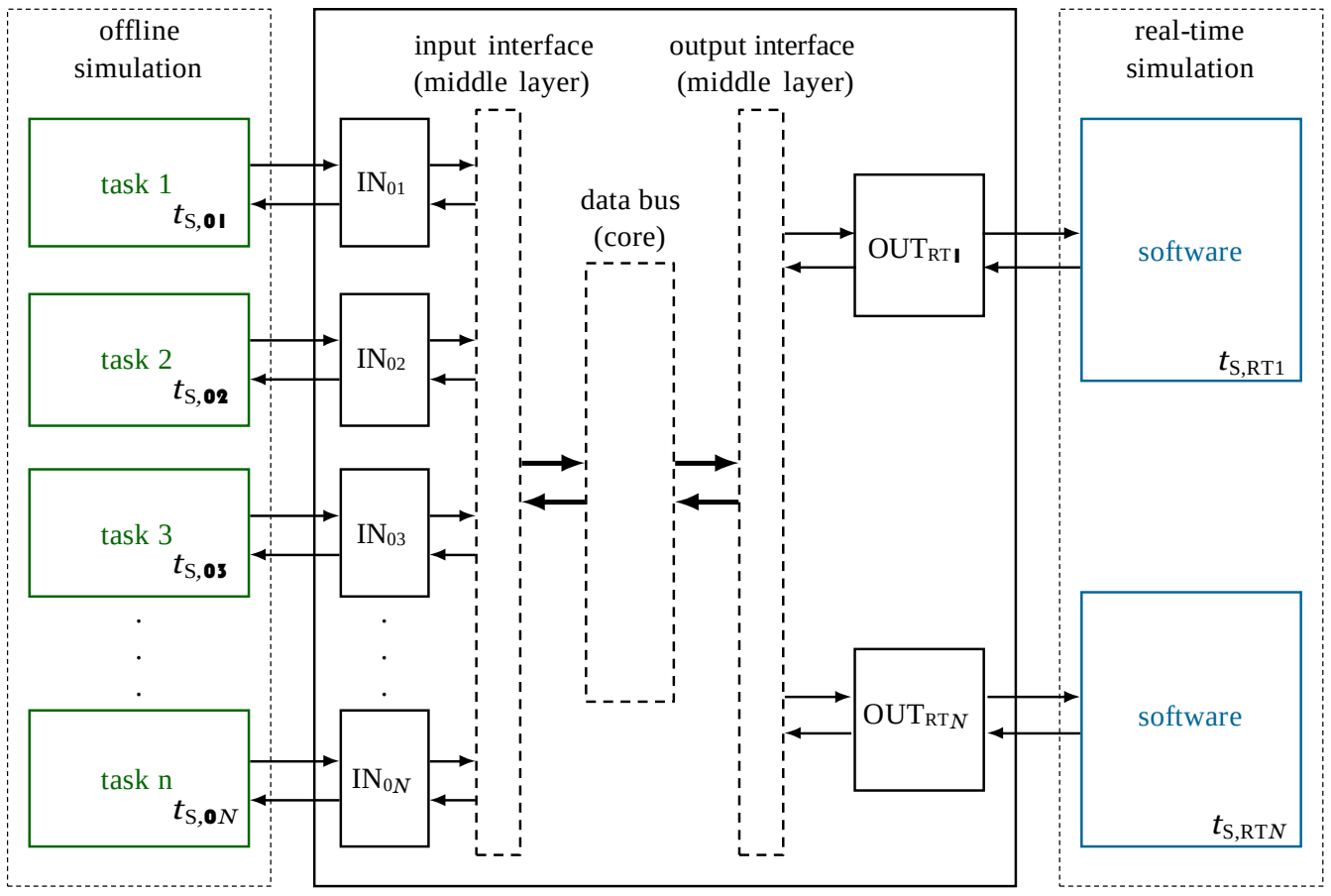

Fig. 1. Simulation message bus architecture for co-simulation of real-time and non real-time systems.

The SMB has been adopted in the development of several co-simulation tools. A first realisation of the ideas in [11] resulted into two different frameworks, i.e., the Lablink, see [12], and the OpSim, see [13]. A comparable approach is VILLAS, see [14], an open source co-simulation platform. While Lablink and OpSim were initially started with the idea of interfacing lab equipment with offline simulation tools, VILLAS was designed to interface geographically distributed research infrastructure and real-time simulators.

## III. Real-time Simulator Integration

In the following, different approaches for the realisation of the SMB concept are introduced and discussed. Also, performance measures are presented.

### A. Lablink Approach

Lablink is a software package which is based on SMB. This kind of communication middleware allows quick and easy coupling of software and hardware components in a lab context. Mainly, Lablink enables different devices that are typically found in an electrical energy laboratory (controllable power sources, controllable loads, grid emulators, measurement devices, etc.) to exchange data and control signals between them, but also with software components such as grid simulators, electric vehicle emulators, etc.

Fig. 2 shows the conceptual structure of Lablink for real-time and non real-time simulations. The left part shows $N$ offline simulation tasks with typical time step size ranges of $t_{S,Oi} \in [100\ ms; 2\ s]$. All tasks are connected to Lablink in an independent and bidirectional way with respect to signal or data exchange. The size range of the time step may heavily vary based on the type of offline simulation, however, typical values are proposed for simulations related to investigations in the electrical domain.

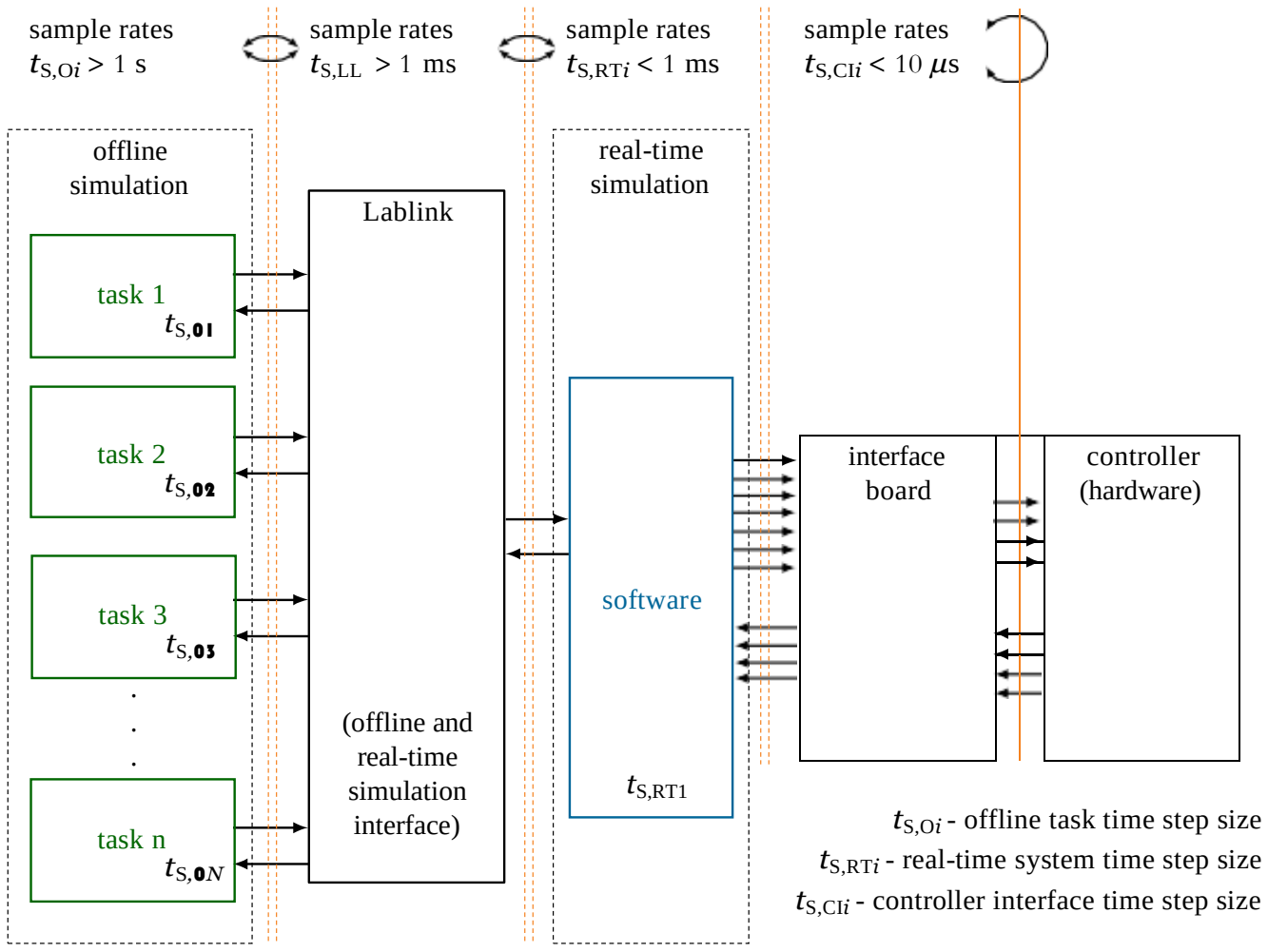

Fig. 2. Lablink structure for real-time and non real-time CHIL applications with indicated sample rates [12].

Lablink is processing and exchanging incoming and outgoing data—as highlighted in the SMB architecture shown in Fig. 1—from offline simulation tasks and from the connected Digital Real-Time Simulator (DRTS), respectively. Hereby, minimum time step sizes of approximately $t_{S,LL} = 1$ ms are specified as operational sample rates for Lablink. The real-time computing system has fixed time step sizes due to the inherent constraint of real-time simulation [15], [16]. For Controller Hardware-in-the-Loop (CHIL) applications, the DRTS typically runs with a time step size in the range of $t_{S,RTi} \in [100\ ns; 1\ ms]$. Sample rates of less than 1 $\mu s$ are required for simulation tasks which aim to emulate Pulse Width Modulation (PWM) signals for control application.

The real-time machine is linked to one or several interface boards, as shown in the right part of Fig. 2. The interfacing boards represent the functional unit between machine controller implemented in hardware and the DRTS system. The number of signals exchanged between the controller and the DRTS may be high. At least, it is higher than the number of signals exchanged between offline tasks and Lablink for typical CHIL applications related to converter control simulations. The maximum specified time step sizes $t_{S,CIi}$ referring to the controller interface are 10 µs. As it has been mentioned above, small time step sizes in the nanosecond range are required when the simulation of PWM signals is explicitly targeted as a use case.

### B. MQTT-based Implementation

The first version of Lablink relies on the Message Queuing Telemetry Transport (MQTT) protocol for the implementation of the data-core. MQTT is a publish-subscribe messaging protocol that relies on a centralised message broker for handling all the data flows.

In order to test the performance of this implementation in terms of signal latency, the scenario described in Fig. 3 was tested. The test consists of 20 offline tasks, that perform a simple *"echo"* function. Each of these tasks were implemented as standalone Python scripts, each running on a single board computer, namely a Raspberry Pi (RPi). More specifically, each task exposes two data signals (one input and one output) via its Lablink Client (LC) and maintains an internal state variable. Every 500 ms each task $i$ increases the value of the internal state variable by one and writes its value (let it be labelled $x$) into the output service, while recording the time-stamp $T_{out}^{i}$ of this event. Other LCs subscribed to this signal will receive the newly output value via the MQTT Lablink data-core. An emq broker running on a dedicated machine connected in the same local area network as the 20 Raspberry Pis was used as the back-end for the MQTT communication needed by the Lablink data-core.

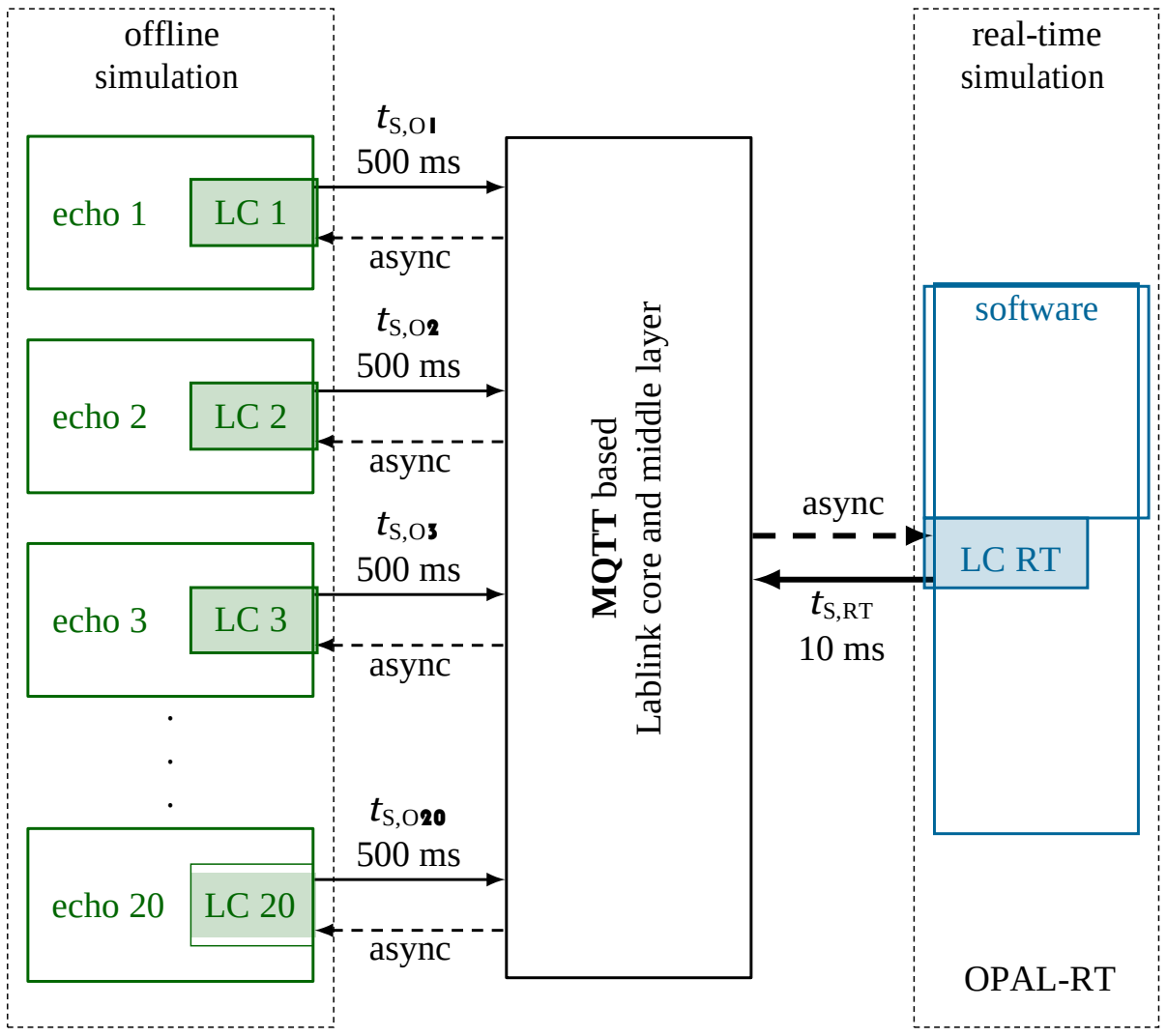

Fig. 3. MQTT-based realisation using SMB as basis.

Meanwhile, each task also monitors its input signals for changes in values. Once the value $x$ is detected, the time-stamp $T_{in}^{i}$ of the event is recorded. The total Round Trip Time (RTT) delay is calculated, as in (1).

$$RTT^{i} = T_{out}^{i} - T_{in}^{i} \tag{1}$$

As can be seen from Fig. 3, OPAL-RT, one of the commercially available DRTS, sits on the other side of the Lablink data-core. A custom LC interfaces the real-time process with the core. The interface between these two processes is done via asynchronous UDP communication. Normally, the real-time process would contain a rather complex or detailed simulation model. However, for this simple benchmarking test the real-time process contains only the UDP communication components and it mirrors back the signals received every 10 ms. In this test OPAL's LC exposes forty signals, i.e., twenty inputs and twenty outputs. The output of each *"echo"* task is connected to a corresponding input of the OPAL client, and vice-versa, in order to create communication loops via the Lablink message bus. The total round trip time delay of these communication loops is calculated, as shown in (1).

In this experiment 1000 signal samples are sent on each communication loop for a total of 20000 samples. The results are displayed in Fig. 4. Here one can see the distribution of the RTT delay for all the messages sent over the Lablink message bus. The minimum RTT delay is 14 ms while the maximum is at 89 ms. Meanwhile, the average RTT delay is at 27.1 ms and in 99% of the cases the delay is smaller than 44 ms. While for compiling these data we used basic *"echo"* tasks, in a real applications these tasks will be complex control algorithms that need to interact with the physical system through sensors and actuators. From this perspective, the presented results imply that it will take on average≈13.5 ms for a sensor present in the physical system, e.g., voltage sensor, to send its data to the external controller.

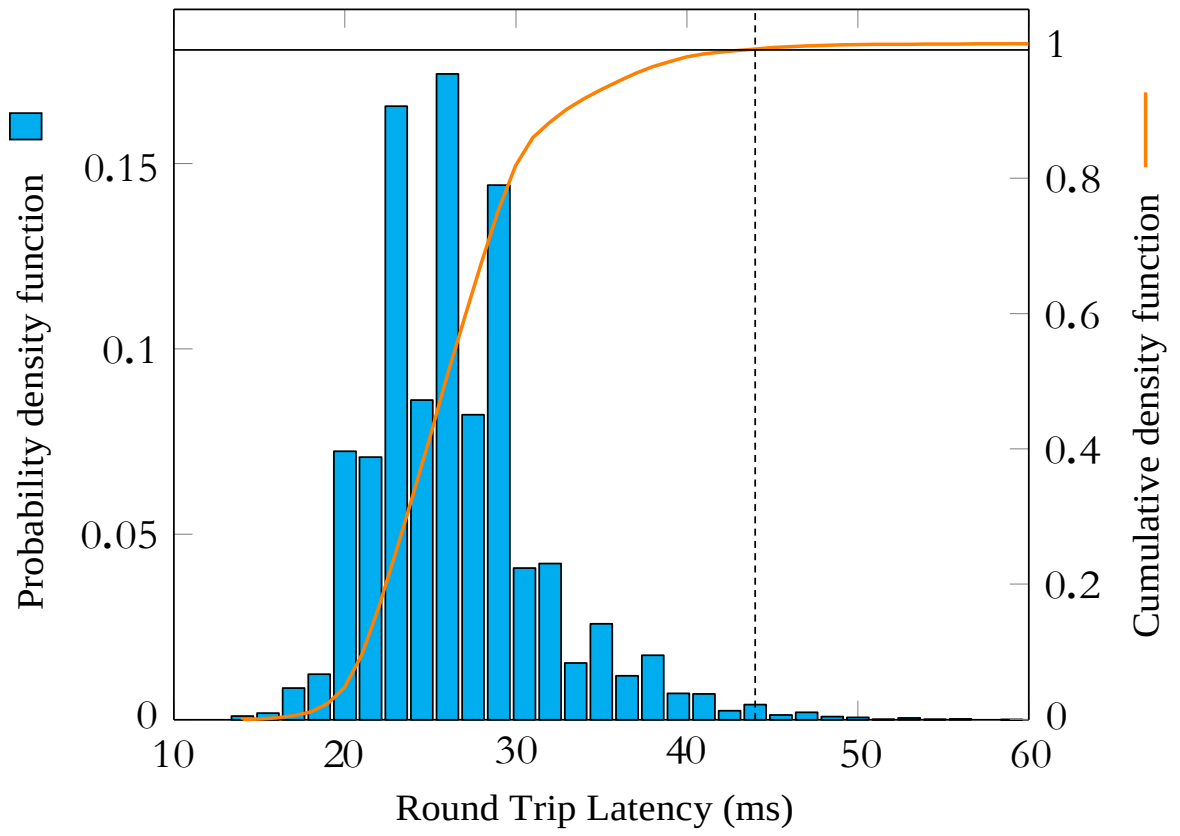

Fig. 4. Round trip latency between the DRTS and the Lablink client. Min: 14 ms. Avg: 27.1 ms. Max: 89 ms. @99%: 44 ms.

### C. OpSim Approach

OpSim is a test and simulation environment with applications ranging from developing prototype controllers to testing operative control software in the smart grid domain. It enables users to connect their software to simulated power systems, or test it in conjunction with other software. The architecture is motivated by the concept in [11], as shown in Fig. 1 and consists of:

- *Message bus:* Components exchange information via a message bus, which forms the center of the OpSim platform and runs as a server application. It uses JavaScript Object Notation (JSON) for its framework and Advanced Message Queuing Protocol (AMQP) with RabbitMQ as message broker.

- *Proxy/Client (P/C) architecture:* Each component in OpSim is situated behind a client and a proxy. The client, implemented in Java, handles the connection, disconnection, synchronisation and information filtering of a component. The proxy is component-specific and "translates" the data between the component and the message bus.
- *Synchronisation:* In [17] the two synchronisation schemes of OpSim are explained. The "real-time" scheme provides synchronisation to a "real" clock which is synchronised with an atomic clock and this scheme is for interaction with external hardware. The "sim-time" scheme introduces a simulation time, which is the base for all components. Further synchronisation is done via this simulated time, in a conservative manner. This "conservative synchronisation" is described in as a way to preserve local causality or to "wait until it is safe".

In [18] the asynchronous integration of an OPAL-RT DRTS to a geographically distributed controller through OpSim and a delay assessment is discussed. OpSim is presented as a black box with a defined functional behaviour, focusing in time-related variables, allowing to analyse the delays between OPAL-RT and OpSim. Although OpSim uses synchronisation schemes for its message bus, the integration of a DRTS into the co-simulation environment is made asynchronously, making the setup asynchronous, as shown in Fig. 5.

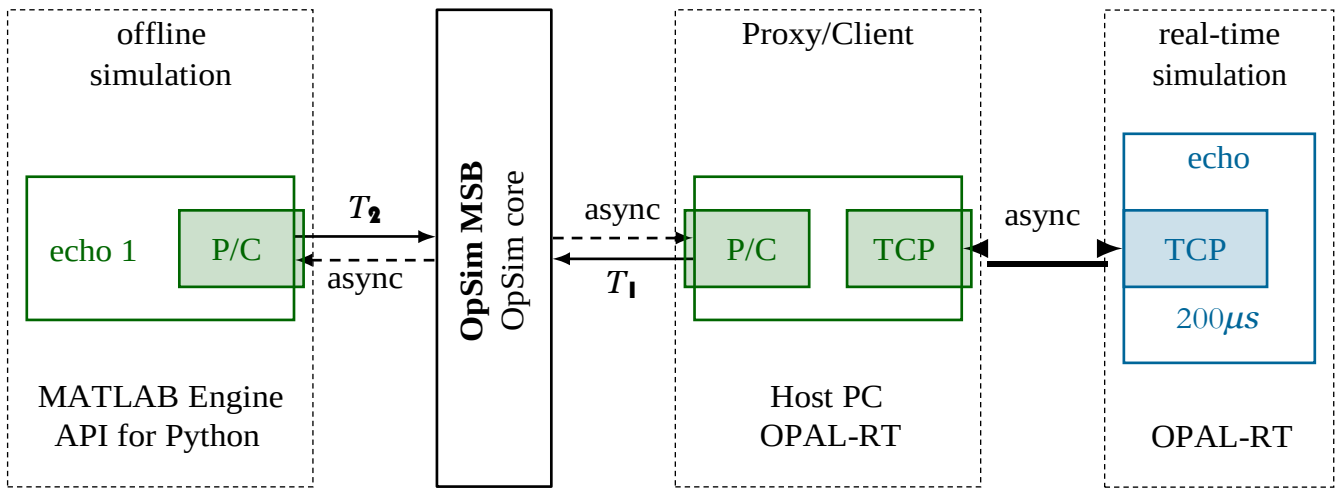

Fig. 5. Integration of OPAL-RT and MATLAB script to OpSim with Java Clients and Python Proxies.

The scenario described in Fig. 5 was used to test the performance of this implementation in terms of signal latency. The test consists of an *"echo"* function for a time variable—local timestamp—being $T_1$ as fast as possible and the OPAL-RT subscribed to its own published signal. The timestamp aims to measure the delays between the OPAL-RT machine and OpSim. The time variable is generated in OPAL-RT and stored as $T^i_{\text{out}}$. To determine the round trip time, the OPAL-RT machine compares each time receives a new value $T^i_{\text{in}}$ using the equation 1. 6563 samples where analysed and the results are displayed in Fig. 6. Here one can see the distribution of the RTT delay for all the messages sent over OpSim. The minimum RTT delay is 14.15 ms while the maximum is at 113.64 ms. Meanwhile, the average RTT delay is at 15.93 ms and in 99% of the cases the delay is smaller than 33.06 ms.

## D. Peer-to-Peer Approach

Both OPSim and Lablink present a centralised implementation of the SMB architecture. While this approach is very flexible and easy to extend, it has the drawback of decreased performance in terms of communication latency, as every signal needs to pass through an intermediate entity, i.e., the data-core.

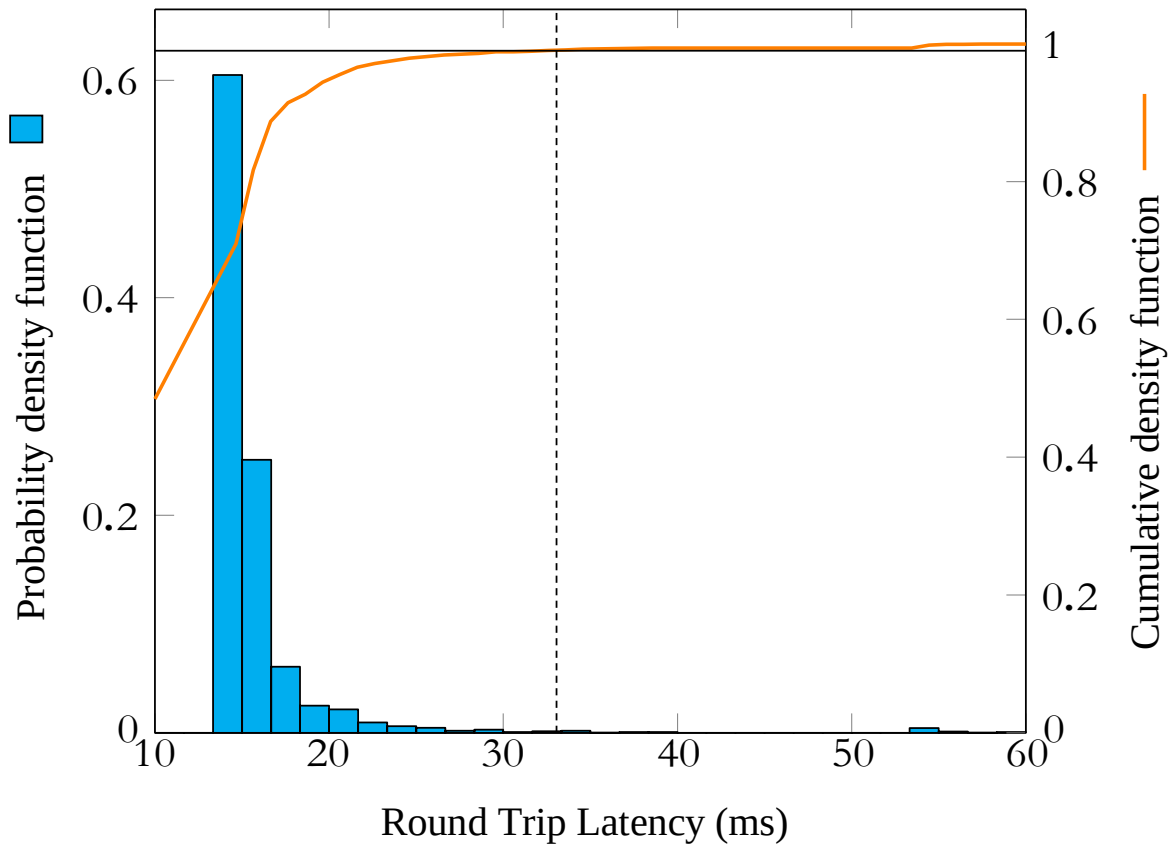

Fig. 6. Round trip latency between the real-time simulator and OpSim MB. Min: 14.15 ms. Avg: 15.93 ms. Max: 113.64 ms. @99%: 33.06 ms.

A complementary solution to these approaches is to employ a decentralised architecture, as shown in Fig. 7. Here, both the offline tasks and the DRTS contain a data-client and server, respectively. Clients can obtain required simulation signals by querying the corresponding server. This creates a direct peer-to-peer connection, thus avoiding an intermediate message bus.

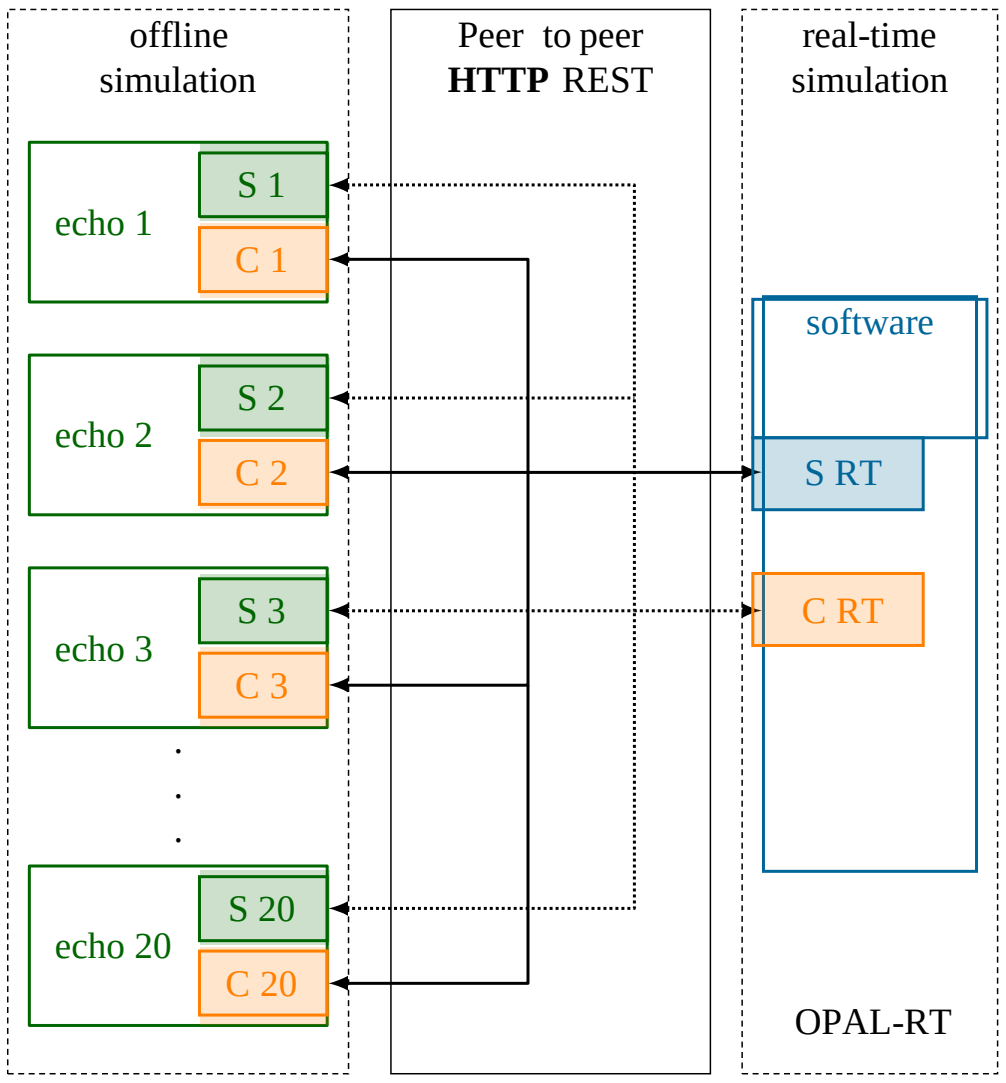

Fig. 7. Peer-to-peer implementation using HTTP REST.

Several options exists for implementing such an approach. Fig. 8 shows the results obtained by using one popular architecture used for addressing interoperability between internet connected devices, namely Representational State Transfer (REST) based on HTTP.

As can be seen from Fig. 8, the round trip signal latency between the offline simulation and the DRTS has an average of ≈6ms, significantly smaller than in the previous approaches that relied on a centralised message bus. However, the improvement in the latency comes with the disadvantage of a less flexible system that requires more configuration and development, and thus is more error prone.

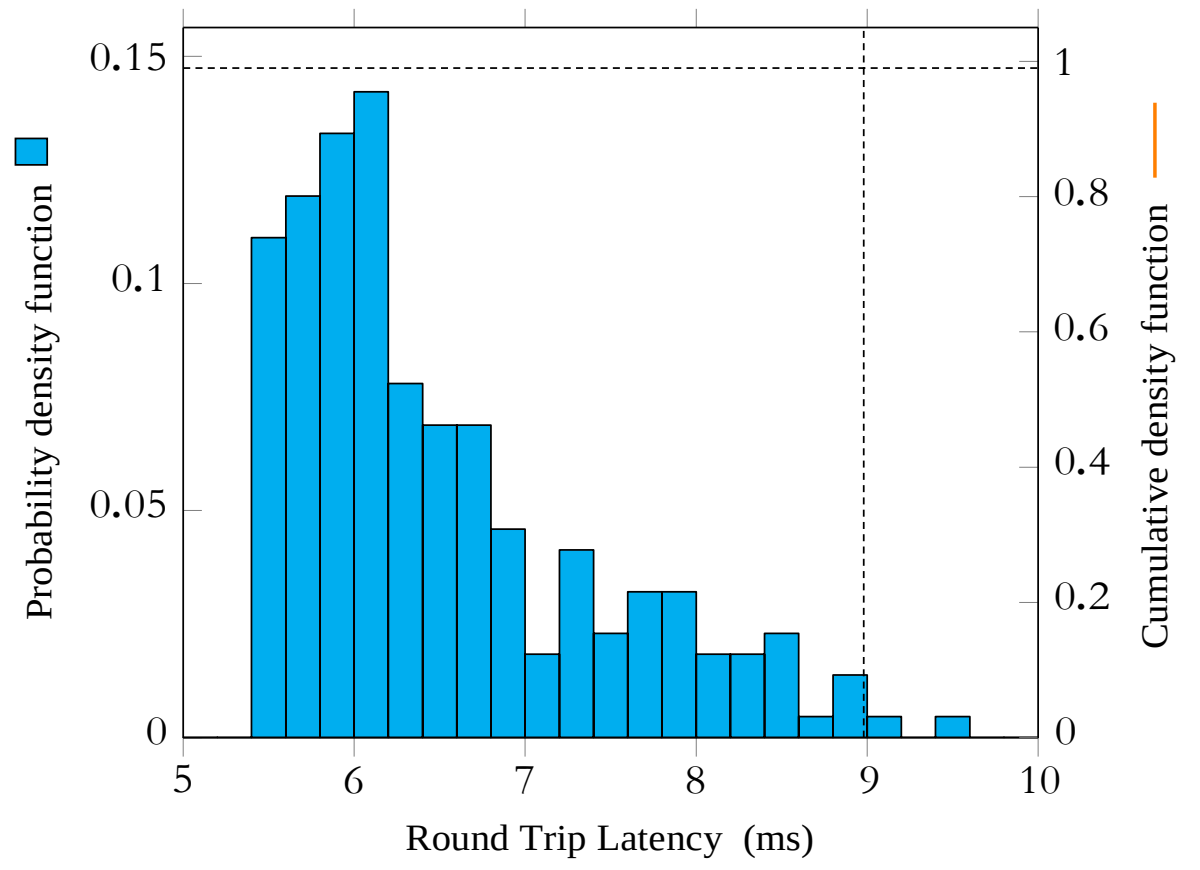

Fig. 8. Round trip latency between the server of the DRTS and the clients of offline tasks. Min: 5.4 ms. Avg: 6.48 ms. Max: 9.48 ms. @99%: 8.98 ms.

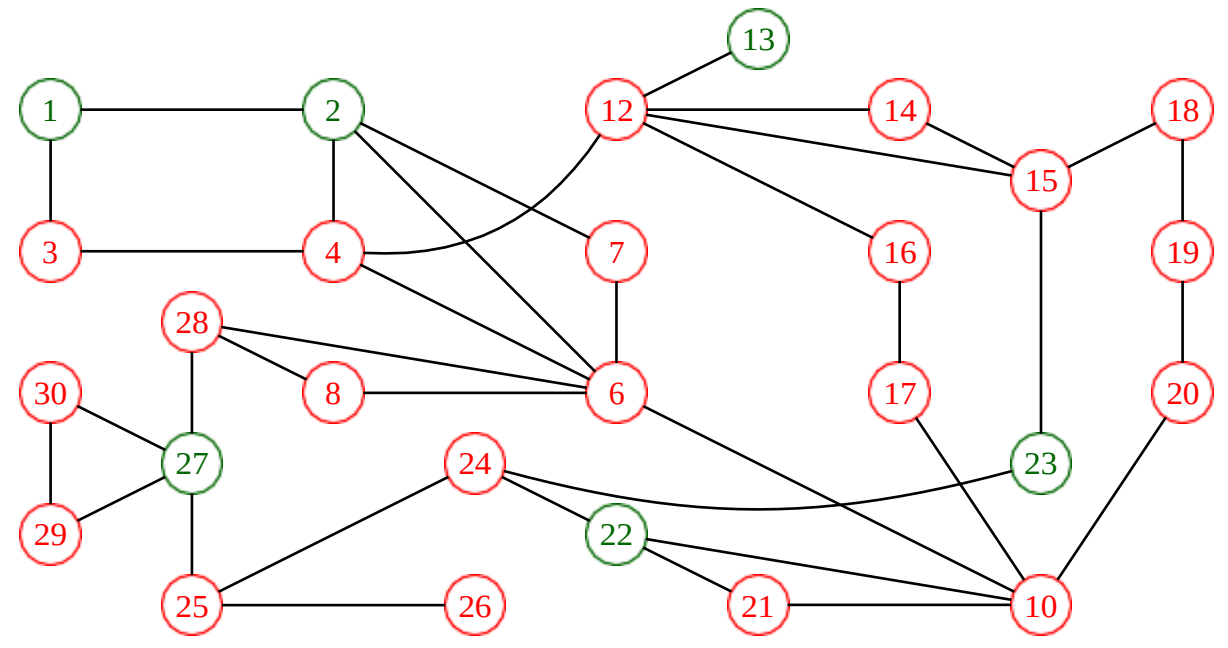

Fig. 9. Graph representation of the 27 terminals MTDC system study case. Red nodes represent loads, while the green ones represent generators.

As both MQTT and HTTP are application layer protocols, in all three implementations the RTT could probably be improved by opting for a lower level protocol such as TCP or UDP for the implementation of the simulation message bus.

## IV. Application Showcase

In the following, two applications that exemplify the type of study cases that can be approached using the proposed simulation framework, are discussed.

### A. *Distributed Optimal Power Flow Control of DC Grids*

In the first application we use the the approach presented in Section III-D for the validation of a distributed multi-agent Optimal Power Flow (OPF) algorithm used as a secondary control layer for dc networks. Fig. 9 shows the graph representation of the network under study. The green nodes correspond to generators, meanwhile the red ones correspond to loads. The operation of each node is controlled by an agent which has complete access to local measurements and, moreover, it is able to communicate with its neighbours. For the inter-agent communication graph the same structure as the one of the electrical network is considered, i.e., if two nodes are electrically connected then the agents in charge are also able to exchange information between them.

Unlike traditional power systems in which the secondary control is centralised, in this case, the agent network forms a distributed control layer. Moreover, it implements a distributed optimization algorithm based on the Alternating Direction of Multiplier Method (ADMM). The full details of the algorithm and how it can be adapted to solve the dc OPF problem can be found in [6].

In brief, ADMM involves an iterative process that requires the agents to execute three distinct steps at every iteration: *(i)* each agent optimises its local state, *(ii)* the results obtained in the previous step are exchanged with the neighbours, and *(iii)* each agent corrects its state based on the information received from the neighbours.

Due to the number of iterations required for convergence, ADMM is often labelled in the literature as a slow algorithm. However, at the moment, optimisation routines in power systems are performed completely offline, and even basic secondary control actions involve time-frames in the range of minutes. Taking this into consideration and given the performance of modern communication protocols, it is interesting to investigate the feasibility of ADMM in this context. The proposed peer-to-peer message bus framework in conjunction with a communication network emulator is used in this example in order to obtain an idea of what time-frames are to be expected from a distributed OPF approach based on ADMM.

The hardware used for this scenario was OPAL-RT and 27 single board RPi computers, all connected to the same dedicated local area network. From the software perspective the local controller or agent in charge of each node was deployed as a Python script to its corresponding RPi. The clients and servers of each agent, as well as the client and server of the OPAL-RT interface, were coupled together according to the data exchange requirements of the use case, e.g., all controllers need to be able to read data from the OPAL-RT, controller 1 needs to exchange information with controllers 2 and 3, etc. The emulation of the communication infrastructure was achieved by using the netem (network emulation) and tc (traffic control) UNIX utilities at the level of each network card connected to the system.

With this setup we were able to perform several experiments using different parameters for the quality of service of the communication. According to the performed experiments, if the distributed control agents are connected over a dedicated local area network, 300 ADMM iterations require 62 s to converge. If 3G communication is used, the convergence time increases to 305 s. Meanwhile, 4G communication yields a convergence time of 167 s.

### B. *Coordinated Voltage Control*

The integration of a DRTS into the OpSim co-simulation was made in [18], to perform a geographically distributed

CHIL experiment between a Coordinated Voltage Control (CVC) script in MATLAB, located in Athens, Greece, and a low-voltage feeder model, based on the CIGRE low-voltage benchmark model, implemented in an OPAL-RT OP5600 simulator located in Kassel, Germany, as shown in Fig. 10.

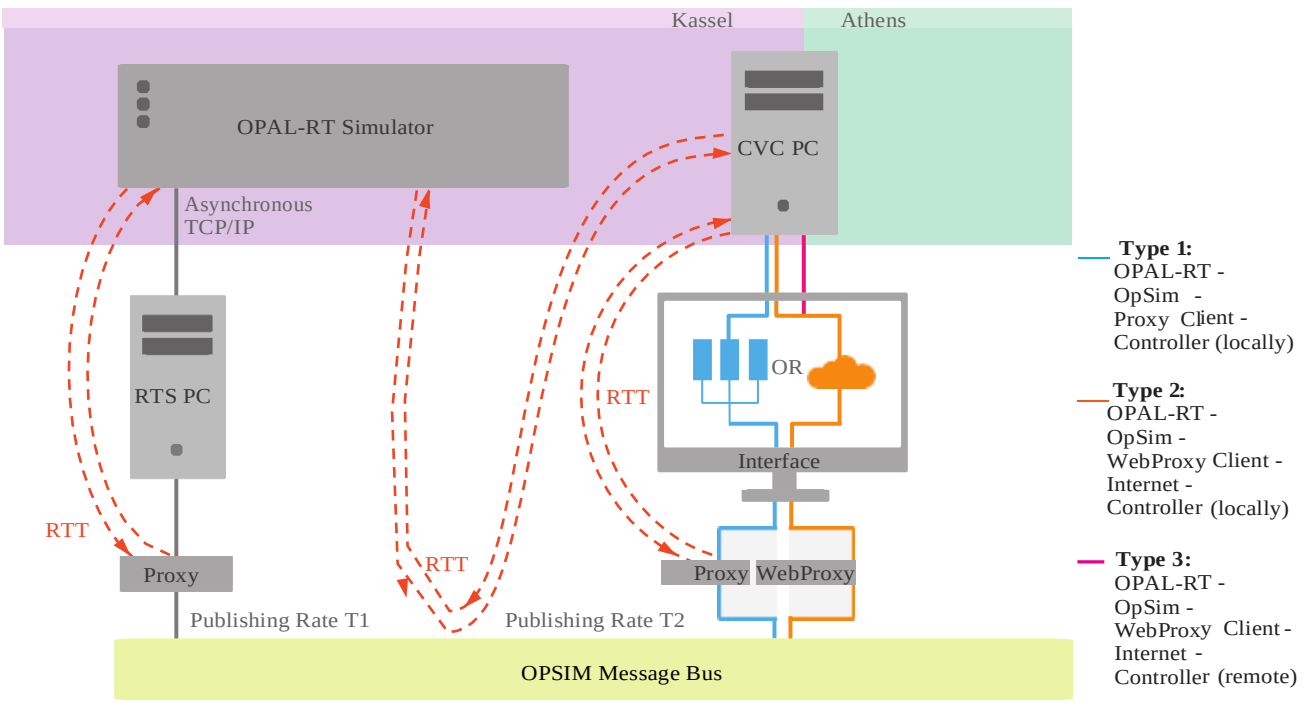

Fig. 10. Test setup for geographically distributed co-simulation with OpSim

The CVC is responsible for monitoring and controlling the state of the low-voltage feeder, by optimising the operation of all voltage-regulating devices present in the network (OLTC, BESS, PVs, etc.). The CVC algorithm's main function is the solution of an OPF problem. A detailed description of the CVC algorithm can be found in [19].

Prior to execute the CHIL experiment, a delay assessment between the CVC PC and OpSim was performed, the same way as presented in Section III-C for Fig. 5. The results of the assessment are presented with the $99^{th}$ percentile for each type described in the legend of Fig. 10, where the RTT for Type 1 is less than 66.8 ms, for Type 2 less than 55.58 ms and for Type 3 less than 128 ms. For the three types, the proxy implemented in Python and the MATLAB Engine API for Python adds to their measurements at least 190.1 ms of delays. Also, the solution of the OPF is always changing, with peaks up to 3 s.

For the co-simulation test case, the experiment was performed multiple times, using different publish rates for the OpSim MB, taking care of the latencies measured in the delay assessment and defined boundaries. A publish rate of 5s was selected for the CVC controller, while the OPAL-RT simulator was tested with 5 different publish rates: 500 ms, 1 s, 2 s, 3 s, and 4 s. As the deviations in the results among the different publish rate experiments are small, so, for simplicity, only the results of the 500ms publish rate experiment were presented in [18].

A similar set-up was also realised locally without using any Internet connection with the Lablink and comparable results have been achieved.

## V. Conclusions

Proper validation and testing methods for analysing smart grid solutions and applications addressing the system-level are necessary today. In this paper an asynchronous integration approach and corresponding implementations have been introduced and discussed.

As it can be seen from the above discussion, coupling a DRTS with other software components introduces a signal latency of at least a few ms depending on the used approach. With some engineering effort, this latency might be further reduced, for example, by using lower-level protocols or faster networks for the communication. However, a considerable improvement below the 1 ms threshold cannot be expected. This will unavoidably limit the scope of application of such a framework to scenarios with slower dynamics, or ones that present a considerable time-scale separation between the different components.

Nevertheless, while not suitable for fast transient analysis, even with the performance of the implementations presented in this paper, there are a vast range of smart grid applications that can be emulated using such a framework. For example, applications similar to the distributed OPF control, where complexity arises from the interactions between a large number of networked components.

The great advantage of the combined DRTS and SMB approach is its flexibility and scalability. The possibility of feeding real-time data to various emulators or domain-specific tools, increases the number and type of scenarios that can be approached with such a system. A good example is the ERA-Net Smart Grids Plus LarGo! project that proposes to investigate the problem of large scale software roll-out in a future software-dominated electric grid. In the scenarios proposed in the project, a large number of models and components from different domains, such as building energy management systems, smart secondary substations, electrical distribution networks, communication networks equipped with digital safety and security features, as well as software roll-out platforms and strategies, all need to be orchestrated together. In order to achieve this, LarGo! uses an approach similar to the ones presented in this paper for the lab validation.

## Acknowledgment

This work is supported by the European Community's Horizon 2020 Program (H2020/2014-2020) under project "ERIGrid" (Grant Agreement No. 654113). Further information is available at the corresponding project website erigrid.eu.